\documentclass[aps,prapplied,twocolumn,superscriptaddress,longbibliography]{revtex4-2}

\usepackage{dcolumn}
\usepackage{bm}
\usepackage{xcolor}
\usepackage{xr-hyper}
\usepackage{amsmath,amsfonts,amssymb}
\usepackage{graphicx}
\usepackage{siunitx}
\usepackage{natmove}
\usepackage{hyperref}

\newcommand{\Eqref}[1]{Eq.~(\ref{#1})}

\newcommand{\norm}[1]{\left|{#1}\right|}
\newcommand{\vect}[1]{\boldsymbol{\mathbf{#1}}}
\newcommand{\univect}[1]{\hat{\boldsymbol{\mathbf{#1}}}}

\begin{document}


\title{Optimal design of nanomagnets for on-chip field gradients}

%
\author{W.~Legrand}
\thanks{These two authors contributed equally}
\affiliation{C12 Quantum Electronics, Paris, France}

\author{S.~Lopes}
\thanks{These two authors contributed equally}
\affiliation{C12 Quantum Electronics, Paris, France}
\affiliation{Universit\'{e} de Lorraine, Institut Jean Lamour, UMR CNRS 7198, Nancy 54011, France}

\author{Q.~Schaeverbeke}
\affiliation{C12 Quantum Electronics, Paris, France}

\author{F.~Montaigne}
\affiliation{Universit\'{e} de Lorraine, Institut Jean Lamour, UMR CNRS 7198, Nancy 54011, France}

\author{M.~M.~Desjardins}
\affiliation{C12 Quantum Electronics, Paris, France}
\date{\today}

%
\begin{abstract}
The generation of localized magnetic field gradients by on-chip nanomagnets is important for a variety of technological applications, in particular for spin qubits. To advance beyond the empirical design of these nanomagnets, we propose a systematic and general approach based on the micromagnetic formulation of an optimal field gradient source. We study the different field configurations that can be realized and find out quantitatively the most suitable ferromagnetic layer geometries. Using micromagnetic simulations, we then investigate the minimum requirements for reaching magnetic saturation in these nanomagnets. In terms of either longitudinal or transverse field gradient, the results provide an optimal solution for uniform, saturated nanomagnets, where the magnetic material can be selected according to the strength of the external fields that can be used.
\end{abstract}
\maketitle
\section{Introduction}\label{sec:introduction}
%

%
On-chip lithographed nanomagnets find a demanding application in the thriving field of quantum technologies, where they are used to generate local magnetic field gradients.
By shaping the energy spectrum of the electrons trapped in quantum dots, they allow for addressing spin qubits by microwave radiation, through the process of electric dipole spin resonance (EDSR) \cite{Tokura2006, Pioro-Ladriere2008, Obata2010}. 
A strong and localized field gradient directly determines the coupling rate of a spin qubit to the microwaves, and is thus a key ingredient in several types of device operations \cite{Hu2012, Beaudoin2016, Benito2019}.
Since the first successes of EDSR with nanomagnets, their design has been optimized through iterative improvements to increase the field gradients \cite{Pioro-Ladriere2007, Lachance-Quirion2015, Mi2018, DumoulinStuyck2021, Harvey-Collard2022}.
This enabled high-fidelity (99.9\%) one-qubit gates on spin qubits with EDSR \cite{Yoneda2018}, reaching the threshold enabling successful error correction codes.
A next step is to enable high-fidelity two-qubit gates in devices hybridizing spins with single photons from on-chip microwave resonators \cite{Borjans2019,Benito2019,Harvey-Collard2022}, which requires even stronger field gradients.
However, obtaining a strong inhomogeneity for the divergence-free, stray magnetic fields of the nanomagnets remains experimentally challenging, considering the lengthscale of around \SI{100}{nm} that is relevant to these devices.
Most of previous efforts have consisted in finding a suitable shape (commonly a split pair of aligned and elongated bars) for the patterned magnets to be placed in the vicinity of the quantum dot electrodes, in combination with increasing their thickness as much as possible.
This has enabled significantly inhomogeneous magnetic field profiles at the nanoscale, with field gradient values approaching the order of \SI{1}{\milli\tesla\per\nano\meter}. Between quantum dots typically separated by 60--\SI{120}{\nano\meter}, these gradients translate into an asymmetric component of the field that can reach an about \SI{30}{\milli\tesla} \cite{Takeda2016} or \SI{40}{\milli\tesla} \cite{Harvey-Collard2022} difference along one direction, thereby modifying the local spin projection axis.
One might still wonder whether this approach is optimal in comparison to any possible geometry in general.
Other forms than a split-pair of bar magnets have indeed been proposed and used in several contexts \cite{McNeil2010, Forster2015}. 
Optimal designs for these magnetic elements remain to be identified, even more when the non-uniformity of the internal magnetization is taken into account.
We describe in this article the realization of nanomagnets as close as possible from ideal magnetic field gradient sources, and identify several design rules relevant for practical nanolithographed devices on chip.
Even though we discuss here the optimization of nanomagnets for spin qubits more specifically, these findings could be used in other domains where strong, nanoscale magnetic field gradients are desirable, e.\,g., in spintronics, in topological superconductivity \cite{Zhou2019,Kornich2020,Turcotte2020,Jardine2021}, for magnetic actuation of nanoelements and nanobeads, or for scanning probe near-field magnetic sensing \cite{Longenecker2012}.
In an initial step, we identify the ideal magnetization distributions that maximize a field or a gradient of any component, when no constraints are imposed on the magnetization.
Even though these ideal field distributions are not achievable in devices, they will allow us to deduce the best nanomagnet shapes when the geometrical constraints of nanofabrication are to be taken into account.
These shapes can be adapted, for instance, to the cases of a double quantum dot defined in a two-dimensional electron gas or along a nanowire.
We then consider systems of nanomagnets that are uniformly magnetized under an external field, as well as nanomagnets hosting magnetic domain walls, and compare the achieved field inhomogenity for all combinations of three orthogonal orientations of magnetization and generated gradient.
The choice of the magnetic material composing the nanomagnets, which could differ from the commonly used elemental Co, is another important aspect to be considered.
While a larger saturation magnetization generates increased field gradients in the saturated case, it might not be true anymore when the internal dipolar interactions lead to the formation of inhomogeneous magnetic domains.
We perform micromagnetic simulations to assess the minimal requirements for keeping a uniform magnetization inside the nanomagnets, and thus for ensuring that they efficiently produce the desired field gradients.
This provides, for a given saturation magnetization, a required value for the external field, a value for the obtained field gradients, as well as a figure of merit for the achieved field inhomogeneity.
We notably show that antisymmetric components of the stray fields larger than the symmetric component can be achieved, as well as we predict a field difference of up to \SI{500}{\milli\tesla} between sites for CoFe nanomagnets and nanowire dots separated by up to \SI{400}{\nano\meter}.

%
\section{Ideal field gradient sources}\label{sec:ideal}
Along the following, we note $\vect{B}(\vect{r})$ the three-dimensional stray fields obtained from an ensemble of nanomagnets.
Depending on the specific requirements of the system, it might be desired to optimize, for example, the absolute value of one component of the magnetic field or its inhomogeneity over some region(s) of interest.
We write $Q$ for this value to optimize, typically a single component of $\vect{B}(\vect{r})$ or one of its derivatives.
The stray fields and their partial derivatives can be deduced from the distribution of elementary magnetic dipoles that fills the source nanomagnets, expressed by a volumic magnetization $\boldsymbol{M}(\boldsymbol{r})=M_{\rm{S}}\vect{m}(\boldsymbol{r})$, where $M_{\rm{S}}$ is the saturation magnetization of the magnetic material and $\vect{m}(\boldsymbol{r})$ is the unit vector giving the local orientation of the magnetization.
The optimization of $Q$ can be conveniently performed relying on the additivity of the magnetic stray fields with respect to the source magnetization distribution.
For each point of space, the best orientation of the local elementary dipole maximizing $Q$ can be found.
As long as there is only one linear quantity to optimize (as opposed to a ratio of them, for example), the problem remains linear and the optimal distribution of the magnetization together with the achieved field can be directly reconstructed.
This approach is in particular also valid for the optimization of a field or a gradient difference between two regions, which is the main focus of this work.
%

%
\begin{figure*}
    \centering

        
    

    \includegraphics[clip,width=7in]{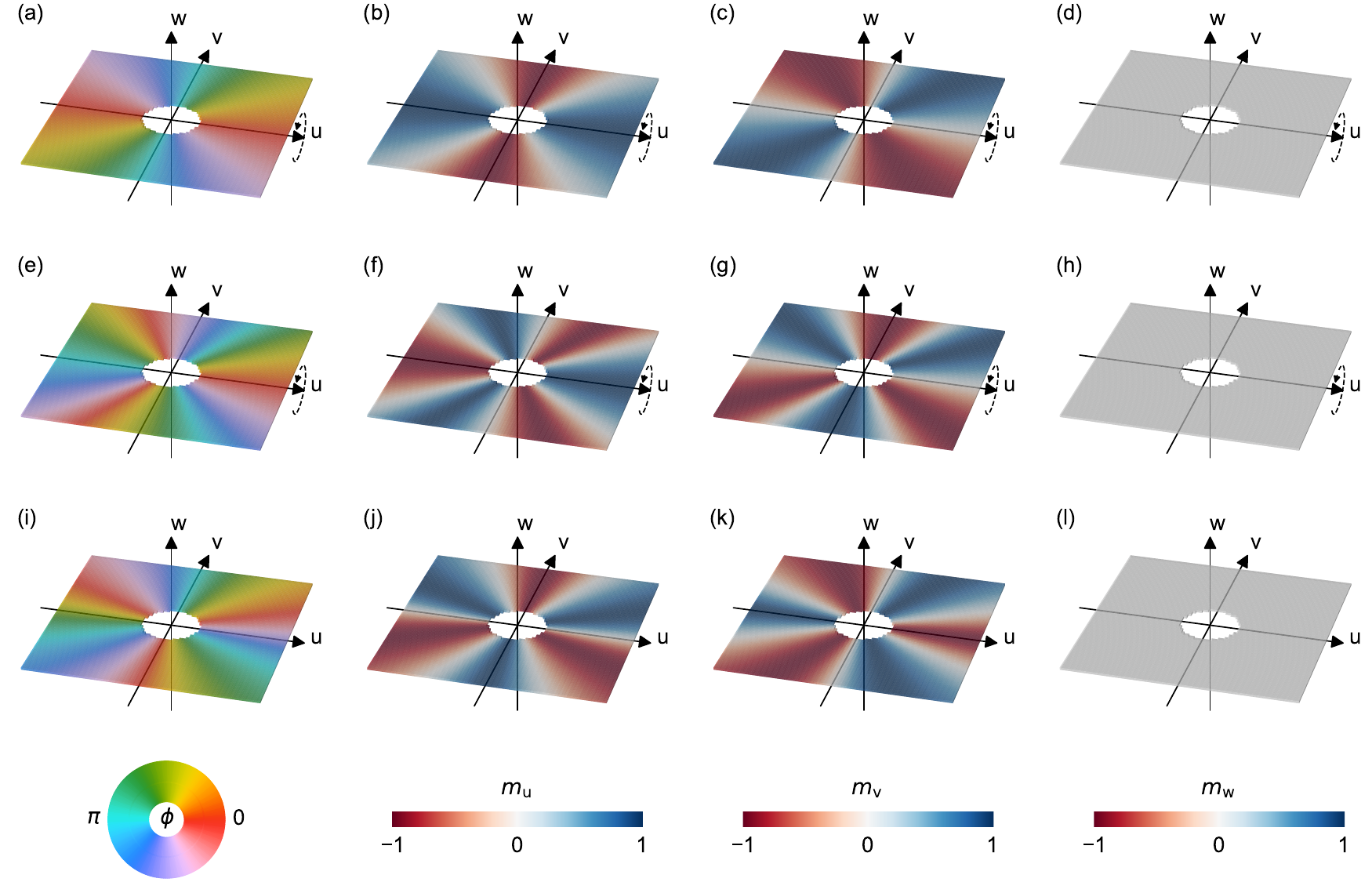}
    \caption{Cut views in the (u,v) plane of the ideal magnetic distributions for optimising (a--d) $B_{u}(0)$, (e--h) $\partial{}B_{u}/\partial{}u(0)$ and (i--l) $\partial{}B_{v}/\partial{u}(0)$. (a,e,i) show the in-plane angle $\phi$, (b,f,j) show $m_{\rm{u}}$, (c,g,k) show $m_{\rm{v}}$, (d,h,l) show  $m_{\rm{w}}$ for these optimal distributions. Axes arrows for $u$, $v$ and $w$ have the same relative length.}
    \label{fig:mideal}
\end{figure*}

We show several examples of optimal distributions of $\vect{m}(\boldsymbol{r})$ in Fig.\ \ref{fig:mideal}.
These differ by their polar profiles around $\boldsymbol{r}=0$ where $Q$ is evaluated, which originate from the form and symmetry of the dipole fields.
Into details, we consider an elementary dipole located at $\vect{r}_{j}$ and a quantity $Q$ to optimize at $\vect{r}_{i}$, their distance $r_{ji}=\norm{\vect{r}_{i}-\vect{r}_{j}}$ and the orientation vector $\vect{e}_{ji}=(\vect{r}_{i}-\vect{r}_{j})/r_{ji}$ (see Supplementary Fig.\ S1 for a schematic description of coordinates), as well as ${\rm{d}}\mu=M_{\rm{S}}{\rm{d}}^3r_{j}$ the magnetic moment contained in the elementary volume at $\vect{r}_{j}$.
Any field component, or its longitudinal and transverse gradients along an arbitrary direction, can be expressed after a choice of matching $u,v,w$ Cartesian coordinates, for which we only express a minimal set of equations.
We start from:
\begin{align}
\label{eq:def_dBijx}
{\rm{d}}\vect{B}_{ji}\cdot\univect{u}&=\frac{\mu_{0}{\rm{d}}\mu}{4\pi{}r_{ji}^3}\left[3(\vect{e}_{ji}\cdot\univect{u})(\vect{e}_{ji}\cdot\vect{m}_{j})-\vect{m}_{j}\cdot\univect{u}\right] \\
\begin{split}
\label{eq:def_ddBijxdx}
{\rm{d}}\frac{\partial\vect{B}_{ji}\cdot\univect{u}}{\partial{}u}&=\frac{\mu_{0}{\rm{d}}\mu}{4\pi{}}\frac{\partial{}r_{ji}^{-3}}{\partial{}u}\left[3(\vect{e}_{ji}\cdot\univect{u})(\vect{e}_{ji}\cdot\vect{m}_{j})-\vect{m}_{j}\cdot\univect{u}\right] \\
&\hphantom{=}+\frac{\mu_{0}{\rm{d}}\mu}{4\pi{}r_{ji}^3}\left[3\frac{\partial}{\partial{}u}(\vect{e}_{ji}\cdot\univect{u})(\vect{e}_{ji}\cdot\vect{m}_{j})\right]
\end{split} \\
\begin{split}
\label{eq:def_ddBijydx}
{\rm{d}}\frac{\partial\vect{B}_{ji}\cdot\univect{v}}{\partial{}u}&=\frac{\mu_{0}{\rm{d}}\mu}{4\pi{}}\frac{\partial{}r_{ji}^{-3}}{\partial{}u}\left[3(\vect{e}_{ji}\cdot\univect{v})(\vect{e}_{ji}\cdot\vect{m}_{j})-\vect{m}_{j}\cdot\univect{v}\right] \\
&\hphantom{=}+\frac{\mu_{0}{\rm{d}}\mu}{4\pi{}r_{ji}^3}\left[3\frac{\partial}{\partial{}u}(\vect{e}_{ji}\cdot\univect{v})(\vect{e}_{ji}\cdot\vect{m}_{j})\right],
\end{split}
\end{align}
where ${\rm{d}}\vect{B}_{ji}$ is the field generated at $\vect{r}_{i}$ by the elementary moment ${\rm{d}}\mu$ at $\vect{r}_{j}$.
Equations \eqref{eq:def_dBijx}--\eqref{eq:def_ddBijydx} describe a field, a longitudinal field gradient and a transverse field gradient optimization, respectively.
It appears from Eqs.\ \eqref{eq:def_dBijx} and \eqref{eq:def_ddBijxdx} that to maximize a field or longitudinal gradient, the local vector $\vect{m}_{j}$ at each $\vect{r}_{j}$ needs to be contained within the plane defined by $\vect{e}_{ji}$ and $\univect{u}$.
Indeed, for a given quantity ${\rm{d}}Q$ obtained with any $\vect{m}_{j}$ within this plane, letting a component of $\vect{m}_{j}$ transverse to this plane reduces all terms by a common factor. 
This property is also easily checked numerically.
For \Eqref{eq:def_ddBijydx} and a transverse gradient, the situation is more complex and leads to $\vect{m}_{j}$ not contained in this plane in the general case.
For the simpler case of Fig.\ \ref{fig:mideal}, however, $\vect{m}_{j}$ is still contained in the $(u,v)$ plane.
Hence, the formulas below apply to any plane for a field or longitudinal gradient and, for simplicity, are restrained to the $(u,v)$ plane for a transverse gradient.
Writing $\phi$ for the in-plane angle between $\univect{u}$ and $\vect{m}_{j}$ and $\theta$ for the in-plane angle between $\univect{u}$ and $\vect{e}_{ji}$, we obtain that:
\begin{align}
{\rm{d}}\vect{B}_{ji}\cdot\univect{u}&=\frac{\mu_{0}{\rm{d}}\mu}{4\pi{}r_{ji}^3}\left[3\cos{\theta}\cos{(\phi-\theta)}-\cos{\phi}\right] \\
\begin{split}
{\rm{d}}\frac{\partial\vect{B}_{ji}\cdot\univect{u}}{\partial{}u}&=-\frac{3\mu_{0}{\rm{d}}\mu}{4\pi{}r_{ji}^4}\left[\cos{(\phi-3\theta)}\vphantom{+\cos^2{\theta}}\right.\\
&\hphantom{=-\frac{3\mu_{0}{\rm{d}}\mu}{4\pi{}r_{ji}^4}\left[\right.}\left.+\cos^2{\theta}\cos{(\phi-\theta)}\right]
\end{split} \\
\begin{split}
{\rm{d}}\frac{\partial\vect{B}_{ji}\cdot\univect{v}}{\partial{}u}&=-\frac{3\mu_{0}{\rm{d}}\mu}{4\pi{}r_{ji}^4}\left[-\sin{(\phi-3\theta)}\vphantom{+\cos{\theta}}\right.\\
&\hphantom{=-\frac{3\mu_{0}{\rm{d}}\mu}{4\pi{}r_{ji}^4}\left[\right.}\left.+\cos{\theta}\sin{\theta}\cos{(\phi-\theta)}\right].
\end{split}
\end{align}
Note that $\theta$ refers here to the polar angle of $\vect{r}_{i}$ defined with respect to $\vect{r}_{j}$. By defining $\theta'=\pi+\theta$, equivalent formulas as a function of $\vect{r}_{j}$ with respect to $\vect{r}_{i}$ can be obtained. To maximize ${\rm{d}}Q={\rm{d}}\vect{B}_{ji}\cdot\univect{u}$,  ${\rm{d}}Q={\rm{d}}\partial\vect{B}_{ji}\cdot\univect{u}/\partial{}u$ or ${\rm{d}}Q={\rm{d}}\partial\vect{B}_{ji}\cdot\univect{v}/\partial{}u$ we find:
%
\begin{align}
\label{eq:alpha_for_field}
\phi&=\arctan{\frac{3\sin{2\theta}}{3\cos{2\theta}+1}} \\
\label{eq:alpha_for_gradx}
\phi&=\arctan{\frac{-5\sin{3\theta}-\sin{\theta}}{-5\cos{3\theta}-3\cos{\theta}}} \\
\label{eq:alpha_for_grady}
\phi&=\arctan{\frac{5\cos{3\theta}-\cos{\theta}}{-5\sin{3\theta}-\sin{\theta}}},
\end{align}
respectively, which are shown in Fig.\ \ref{fig:mideal}. Here, $\arctan$ implies that the quadrant of the angle is chosen by the signs of the numerator and denominator.
This treatment resembles, but is not equivalent, to the original realization of multipole magnetic fields using permanent magnets by Halbach \cite{Halbach1980}.
It is well known, for example, that the component of the field along the dipole direction is positive only within a cone originating from the dipole position, generated around the dipole axis with an angle of $\arccos{(1/\sqrt{3})}\approx$ \SI{54.7}{deg}.
This is contained in the denominator of \Eqref{eq:alpha_for_field}. 
In addition, Eqs.\ \eqref{eq:alpha_for_field}--\eqref{eq:alpha_for_grady} explain why the optimal distributions for maximizing a component and a gradient of the stray fields see $\vect{m}_{j}$ rotating respectively two and three times when looping around the origin at $\vect{r}_{i}$ \cite{Halbach1980}.
For more complex quantities $Q$ that do not add up, a similar approach is possible but would require a non-linear optimization \cite{Abert2017,Turcotte2020}.
These calculations reveal the ideal internal magnetization distributions that nanomagnets should tend to reproduce to optimize the stray fields and gradients they create.
The full, three-dimensional optimal distributions are deduced from Fig.\ \ref{fig:mideal} by rotation of the figure plane around the $u$-axis for the field and longitudinal gradient cases.
However, this rotation symmetry is absent for the transverse gradient case (see the full representations of this distribution in Supplemental Material S2).
From these ideal distributions, it is possible to get an intuition of what would constitute best magnet shapes, as we demonstrate below.
Beyond the shape of the nanomagnets, one other key aspect for field strength optimization is the minimal nanomagnet-to-dot distance.
This appears clearly by considering an exclusion zone around the dot, of radius $r$, inside which no magnet can be present.
This allows for taking into account several kinds of nano-fabrication constraints preventing from reaching infinitely small sizes or strongly overlapping features.
By performing the integration of ${\rm{d}}Q={\rm{d}}\partial\vect{B}_{ji}\cdot\univect{u}/\partial{}u$ (${\rm{d}}Q={\rm{d}}\partial\vect{B}_{ji}\cdot\univect{v}/\partial{}u$) over the remaining space $r_{ji}>r$, we find that $Q\approx3.353\,\mu_0M_{\rm{S}}/r$ ($Q\approx2.832\,\mu_0M_{\rm{S}}/r$). For ${\rm{d}}Q={\rm{d}}\vect{B}_{ji}\cdot\univect{u}$, we have to introduce also a maximum radius $R$, in which case we find $Q\approx1.380 \,\mu_0M_{\rm{S}}\ln{\left(R/r\right)}$.
This sets a fundamental upper limit on the achievable gradients, and shows that for creating a large field difference between the dots, the nanomagnets need to extend as close as possible from the dots.
%

%
\section{Realization with saturated nanomagnets and domain walls}\label{sec:satDW}
To progress further in the description of realistic magnetic field gradient sources, it is required to be more specific on the quantity to optimize.
We let aside the generation of a localized gradient in a single dot \cite{Kawakami2014} and focus on the generation of a field difference between two dots in a double quantum dot (DQD) system \cite{Laird2007,Obata2010,Shin2010,Yoneda2014,Takeda2016,Mi2018,Samkharadze2018,Harvey-Collard2022}, as a similar approach works in both cases.
An electron in the DQD system is confined to within a local area or length by an electrostatic potential.
Considering dots placed symmetrically with respect to the field distribution, the fields in the two dots of the DQD will be composed of both a symmetric and an antisymmetric part.
We may define, for each field component $i=u,v,w$,
\begin{equation}
B_i^{\rm{s,as}}=(B_{i}^{\rm{R}}\pm{}B_{i}^{\rm{L}})/2,
\end{equation}
where $p=\rm{L,R}$ stand for the left and right dot locations.
The effects of the stray fields on the energies in the left and right part of the DQD can then be expressed by
\begin{equation}
\alpha_{\rm{s,as}}=2\mu_{\rm{B}} \int \sqrt{\sum_i{\left(B_i^{\rm{s,as}}(\vect{r})\right)^2}} \norm{\psi_p(\vect{r})}^2 d\vect{r},
\end{equation}
where $\norm{\psi_p(\vect{r})}^2$ is the probability of presence of the electron in dot $p$, the prefactor 2 stands for the Land\'e factor of the electron spin and $\mu_{\rm{B}}$ is the Bohr magneton.
This supposes a spin quantization axis along $B^{\rm{s}}$ or $B^{\rm{as}}$ and to neglect the dots overlap.
Due to the nanomagnet symmetries in some optimized cases, one of the three components of $B_i^{\rm{s,as}}$ is possibly zero; otherwise, the different field components containing a symmetric or an antisymmetric part are combined into $\alpha_{\rm{s}}$ and $\alpha_{\rm{as}}$.
This corresponds to finding the symmetric and antisymmetric field quantization axes for the two dots.
The gradient should thus be seen as an inhomogeneity between two sites of the fields averaged over each dot.
This extension modifies the optimal distributions shown above, due to their convolution with the probability of presence functions $\norm{\psi_p(\vect{r})}^2$.
For simplicity, in \ref{subsec:satDW:sat} we first introduce the general concept by considering two punctual sites as ideal dots. 
We present later, in \ref{subsec:satDW:finite}, the effects of the average described above in the calculation of $Q=B_{i}^{\rm{R}}-B_{i}^{\rm{L}}$.
%

%
\begin{figure*}
    \centering
    
    
    \includegraphics[clip,width=7in]{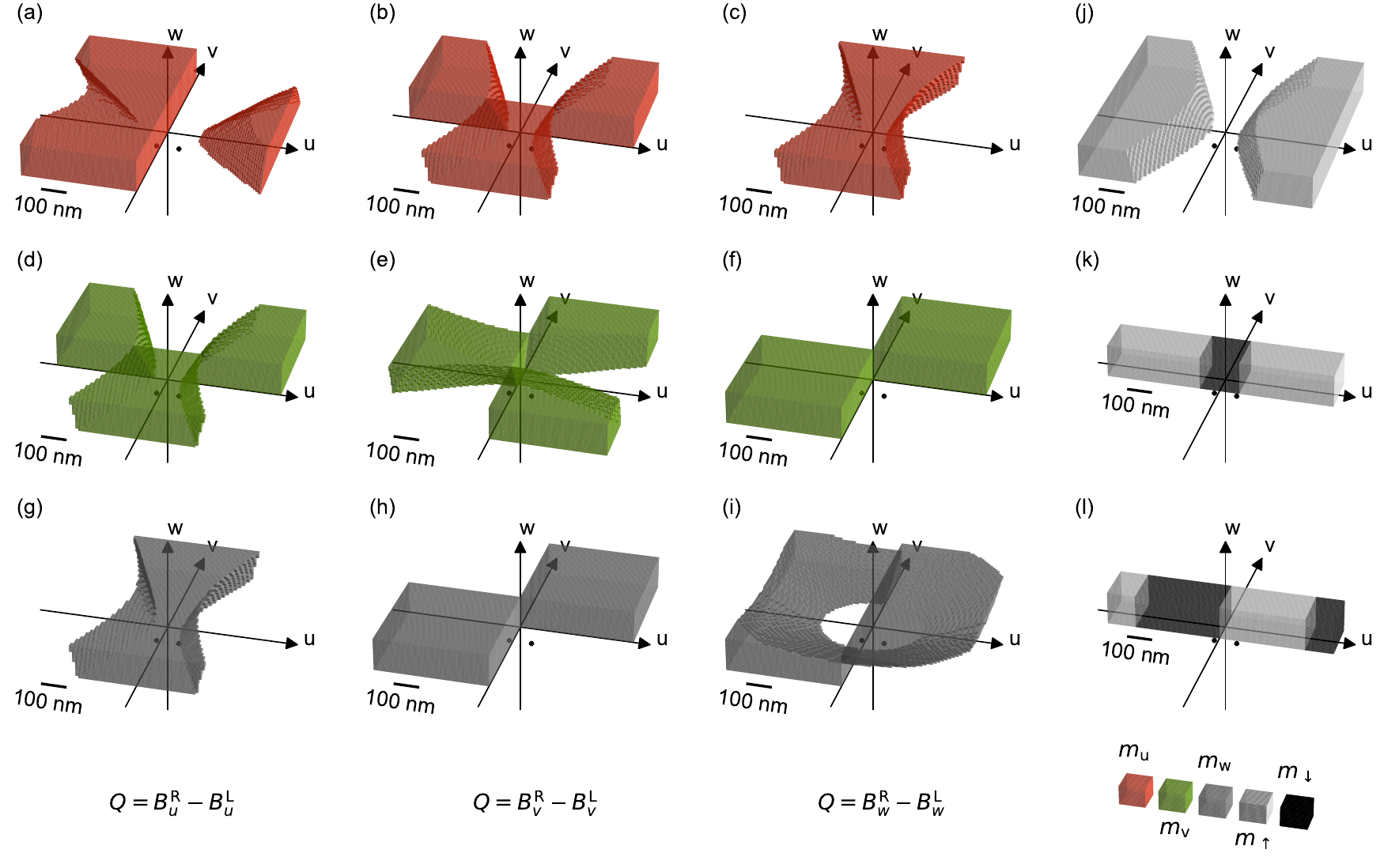}
    \caption{(a--j) Three-dimensional views of optimal, uniformly magnetized nanomagnets generating a field difference in a double quantum dot system. Uniform magnetization imposed along (a--c) $u$, (d--f) $v$ and (g--i) $w$ directions maximizing field difference for (a,d,g) $B_{u}$, (b,e,h) $B_{v}$ and (c,f,i) $B_{w}$. (j) Complementary shape to cases c and g, maximizing a positive field difference of $B_{u}$ and $B_{w}$. (k,l) Domain wall nanomagnets maximizing field difference for (k) $B_{u}$ and (l) $B_{w}$. The light (dark) color filling corresponds to $\vect{m}$ along $w$ ($-w$). Dots locations are indicated by the pair of points. Axes arrows for $u$, $v$ and $w$ have the same length, scale bar equals both the interdot distance and dots depth of \SI{100}{nm}.}
    \label{fig:munif}
\end{figure*}

Most of the time, it will not be possible to impose the distribution of dipoles within the nanomagnets as finely as in Fig.\ \ref{fig:mideal}, for two main reasons: the need for an external magnetic field to be applied to the devices, and the disturbance from internal dipolar interactions.
Both will favor magnetic configurations inside the nanomagnets that are different from the ideal ones for field and gradient generation. 
It is also necessary to consider the constraint of a planar film geometry, due to available deposition techniques and integration of the nanomagnets in the device layout.
We may consider an in-plane magnetization (due to the shape anisotropy of planar films, added to the requirement of employing in-plane external magnetic fields to ensure compatibility with standard superconducting microwave resonators).
Another avenue is to use magnetic materials with strong out-of-plane magnetic anisotropy that do not require out-of-plane external magnetic fields during operation.
With these constraints in mind, a relevant design approach to maximize the field difference between the two dots is to consider uniformly magnetized nanomagnets restrained to a finite thickness.
Their uniform magnetization can be easily ensured by saturation under a large external magnetic field in the in-plane case, or a large magnetic anisotropy in the out-of-plane case, which are both able to overcome internal magnetostatic fields.
In this approach, the optimization simply consists in deciding at each point of space whether part of a magnet should be placed there.
This choice naturally depends on whether the corresponding elementary dipole increases $Q$, given its externally imposed magnetization $\vect{m}$.
In turn, this determines the optimal shapes for saturated nanomagnets.
\subsection{Punctual dots}\label{subsec:satDW:sat}
We first consider ideally punctual dots, such that $\norm{\psi_p(\vect{r})}^2$ is non-zero only at two points L and R and $Q=B_{i}^{\rm{R}}-B_{i}^{\rm{L}}$.
The shapes generating the strongest field differences within such an ideal DQD system are shown in Fig.\ \ref{fig:munif}(a--i), for the $u,v,w$ magnetic orientations and a difference of $B_{u}$, $B_{v}$, $B_{w}$ fields. 
This example consists of a typical DQD system where the dots are separated by \SI{100}{nm} and buried \SI{100}{nm} below the magnetic layer of thickness \SI{200}{nm} across $w$.
Results for other values of these parameters are provided in Supplemental Material S3.
The shapes appearing in Fig.\ \ref{fig:munif}(a--i) are explained by the predictions of the ideal case presented above.
Considering the ideal magnetic distributions $\vect{m}^{\rm{opt}}$ from Fig.\ \ref{fig:mideal}, the suitable shape for a uniformly magnetized magnet with $m_{u,v,w}=1$ can be qualitatively found, with satisfying accuracy, by taking the region of space for which $\vect{u,v,w}\cdot{}\vect{m}^{\rm{opt}}>0$.

\begin{table*}
\caption{\label{tab:munif_vals} Estimated field components in the DQD system for the configurations of Fig.\ \ref{fig:munif}.} 
\begin{ruledtabular}
\begin{tabular}{l@{\extracolsep{\fill}}
    cc
    D{.}{.}{1.3}
    D{.}{.}{1.3}
    D{.}{.}{1.3}
    D{.}{.}{1.3}
    D{.}{.}{1.3}
    D{.}{.}{1.3}
    D{.}{.}{1.3}
    D{.}{.}{1.3}}
 Fig. \ref{fig:munif} &
 $m$ dir. & $\alpha_{as}$ dir. &
 \multicolumn{1}{c}{$B_{u}^{L}$} &
 \multicolumn{1}{c}{$B_{v}^{L}$} &
 \multicolumn{1}{c}{$B_{w}^{L}$} &
 \multicolumn{1}{c}{$B_{u}^{R}$} &
 \multicolumn{1}{c}{$B_{v}^{R}$} &
 \multicolumn{1}{c}{$B_{w}^{R}$} &
 \multicolumn{1}{c}{$\quad\alpha_{\rm{s}}/\mu_{\rm{B}}\quad$} &
 \multicolumn{1}{c}{$\alpha_{\rm{as}}/\mu_{\rm{B}}$} \\
 panel                 &           &                     & \multicolumn{6}{c}{($\mu_0M_{\rm{S}}$)} & \multicolumn{2}{c}{($\mu_0M_{\rm{S}}$)} \\
\hline
 (a) & $u$ & ${u}$ & -0.097 &  0.000 & -0.075 &  0.022 &  0.000 & -0.081 &  0.174 &  0.119 \\
 (b) & $u$ & ${v}$ & -0.037 & -0.035 &  0.003 & -0.037 &  0.035 & -0.003 &  0.075 &  0.071 \\
 (c) & $u$ & ${w}$ & -0.137 &  0.000 &  0.064 & -0.137 &  0.000 & -0.064 &  0.274 &  0.129 \\
 (d) & $v$ & ${u}$ & -0.035 & -0.037 & -0.038 &  0.035 & -0.037 & -0.038 &  0.106 &  0.070 \\
 (e) & $v$ & ${v}$ & -0.000 & -0.073 & -0.000 &  0.000 & -0.001 &  0.000 &  0.074 &  0.072 \\
 (f) & $v$ & ${w}$ &  0.092 & -0.037 & -0.047 &  0.092 & -0.037 &  0.047 &  0.198 &  0.094 \\
 (g) & $w$ & ${u}$ &  0.064 & -0.000 &  0.133 & -0.064 & -0.000 &  0.133 &  0.266 &  0.129 \\
 (h) & $w$ & ${v}$ &  0.003 & -0.047 &  0.074 & -0.003 &  0.047 &  0.074 &  0.149 &  0.094 \\
 (i) & $w$ & ${w}$ &  0.088 & -0.000 &  0.016 &  0.082 & -0.000 &  0.132 &  0.226 &  0.116 \\
 (j) & $u$ & ${w}$ &  0.062 &  0.000 & -0.058 &  0.062 &  0.000 &  0.058 &  0.124 &  0.116 \\
 (j) & $w$ & ${u}$ & -0.058 &  0.000 &  0.016 &  0.058 & -0.000 &  0.016 &  0.032 &  0.116 \\
 (k) & $w$,$-w$ & ${u}$ & -0.078 &  0.000 & -0.058 &  0.078 &  0.000 & -0.058 &  0.116 &  0.156 \\
 (l) & $w$,$-w$ & ${w}$ &  0.128 &  0.000 & -0.088 &  0.128 & -0.000 &  0.088 &  0.256 &  0.176 \\
\end{tabular}
\end{ruledtabular}
\end{table*}

In two cases, for a uniform magnetization along $\vect{u}$ generating a $B_{w}$ field difference [Fig.\ \ref{fig:munif}(c)] and for a uniform magnetization along $\vect{w}$ generating a $B_{u}$ field difference [Fig.\ \ref{fig:munif}(g)], the nanomagnets have a lower symmetry. 
Their complementary shape, represented in Fig.\ \ref{fig:munif}(j), does not generate an equal field difference with opposite sign, but a lower difference instead.
This is due to the finite thickness across $w$ of the nanomagnets and their location above the dots.
It appears overall that two of these shapes are not easily realizable with common deposition and lithography techniques [Fig.\ \ref{fig:munif}(a,i)], due to the requirement of having elevated portions compared to the reference plane $w=0$.
Note that considering three-dimensional shapes of nanofabricated magnets, as can be envisioned from recent promising results in the field \cite{Fernandez-Pacheco2017}, would allow to go beyond this restriction.
Nevertheless, all the other configurations [Fig.\ \ref{fig:munif}(b--h) and (j)] are close to a simple shape that can be obtained from a single and uniformly thick layer, avoiding the use of different thicknesses within one design.
This set of simple configurations covers the needs for generating an asymmetric field component along the three possible directions $u,v,w$ relative to the DQD axis.
The values of the symmetric and antisymmetric field components, obtained for the specific dimensions considered in the present example, are summarized in Table \ref{tab:munif_vals}.
The space discretization has been set to cells of \SI{5}{\nano\meter} to reach the third digit precision in this estimation.
To account fully for cases (c) and (g), we complete the table with the values obtained for their common complementary shape, case (j).
It appears that in-plane transverse gradients [cases (d--f)] are more difficult to achieve, as revealed by their lower values for $\alpha_{\rm{as}}$.
Among the different results, the last case (j) of a longitudinal gradient produced with out-of-plane magnetization is noticeable. 
In comparison to all other configurations, it allows for a much larger asymmetric component relative to the symmetric one, which therefore creates the most inhomogeneous field distribution.
Two patterns among the ones shown in Fig.\ \ref{fig:munif} are expected to be easier to implement in practice, cases (j) and (d). 
This is because the shape anisotropy from the different parts of the magnetic system favors the intended uniform magnetization, as will be shown later.

\subsection{Nanomagnets with domain walls}\label{subsec:satDW:DW}
An additional feature of nanomagnets is the possibility to host a domain wall. 
By allowing for opposite magnetization directions inside different domains, a larger portion of space can be filled with magnetic material, producing in turn a stronger gradient. 
Sharp and well-defined domain walls are easier to obtain in perpendicularly magnetized system, as in magnetic multilayers with interfacial anisotropy, for example.
By introducing a domain wall inside a perpendicularly magnetized nanomagnet, and provided the wall width is narrow compared to the other dimensions in the system, the magnetic system gets closer to an ideal gradient source.
We still consider punctual dots and $Q=B_{i}^{\rm{R}}-B_{i}^{\rm{L}}$.
We present in Fig.\ \ref{fig:munif}(k,l) the optimized nanomagnets with perpendicular magnetization and domain wall(s), now considering a nanomagnet width of \SI{200}{nm} as well.
A constraint was added that the domain wall is straight and contained within a plane perpendicular to $u$, which is the configuration observed in practice as it minimizes the domain-wall energy.
Thus, this optimization consists in choosing a positive or negative $\vect{m}\cdot\vect{w}$ for all positions along $u$.
A domain wall pair is obtained for the case of a $B_{u}$ field difference [Fig.\ \ref{fig:munif}(k)], while three domain walls (a single domain wall would suffice in a shorter nanomagnet) are obtained in the case of a $B_{w}$ field difference [Fig.\ \ref{fig:munif}(l)].
The corresponding field values are reported in Table \ref{tab:munif_vals} as well.
They confirm that a larger field difference is obtained in the end, despite the smaller width of the nanomagnets.

\subsection{Dots with finite extension}\label{subsec:satDW:finite}

%
\begin{figure*}
\centering

    

\includegraphics[clip]{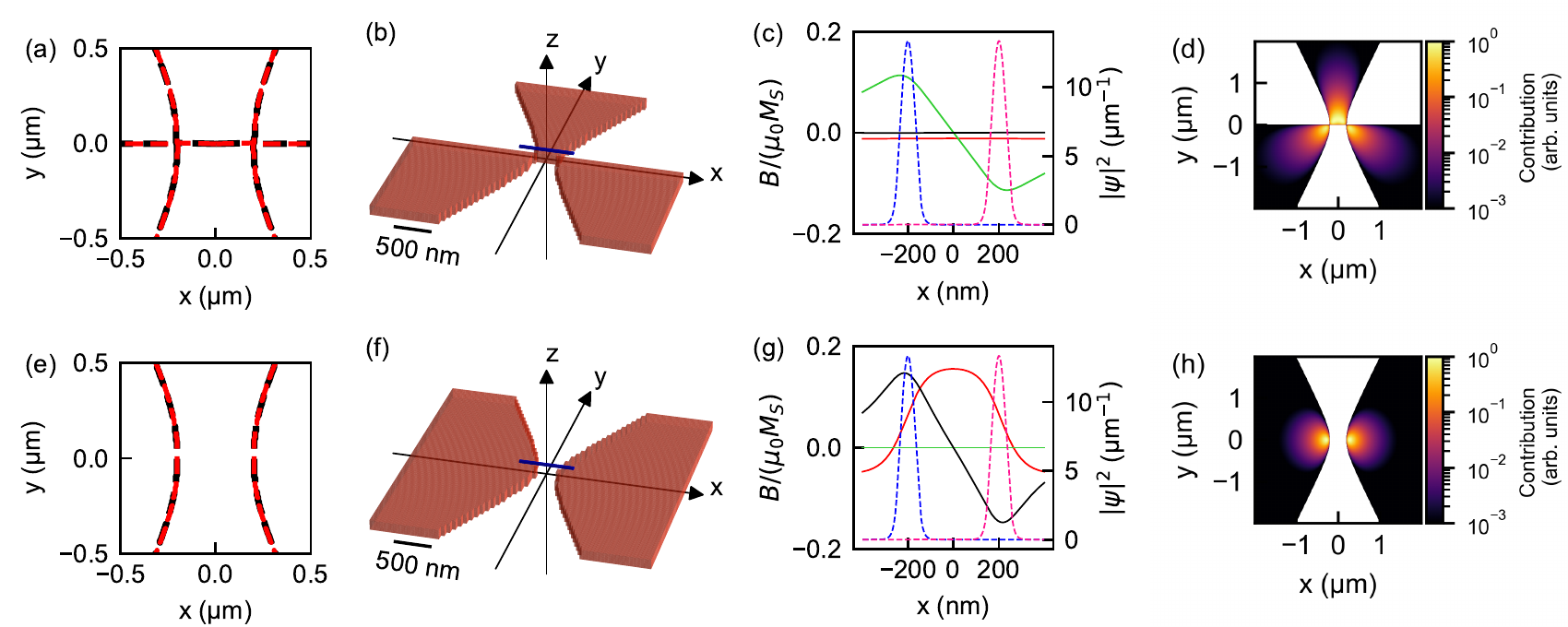}

\caption{Optimal shapes of a saturated nanomagnet in the case of dots of finite extension. Uniform magnetization imposed along $x$, maximizing (a-d) a $B_{y}$ and (e-h) a $B_{z}$ field difference. (a,e) Limits of the optimized nanomagnets for two punctual dots (red lines) or two dots with spatial extension (black lines), and identical locations. (b,f) Three-dimensional views of the nanomagnets obtained for finite-size dots. The nanowire location is indicated by the dark blue line. (c,g) Profile of the stray magnetic fields along the nanowire ($B_{x}$ in red, $B_{y}$ in green and $B_{z}$ in black) and overlap with the probability of presence $\norm{\psi_p(\vect{r})}^2$ of an electron in the left/right dot (respectively in blue/pink). (d,h) Relative contributions, normalized to 1 for the largest value, of each part of the nanomagnet to the maximized quantity $Q$.}\label{fig:idealcase2}
\end{figure*}

We now address into more details the impact of the spatial extension of the individual dots of the DQD system.
A pair of dots with linear confinement is now considered, as found in a nanowire DQD \cite{Viennot2015,Cubaynes2019}.
The dots are aligned along $x$ and defined between \SI{-250}{\nano\meter} and \SI{-150}{\nano\meter} for dot $L$, and between \SI{150}{\nano\meter} and \SI{250}{\nano\meter} for dot $R$. The nanowire is suspended at $z=$ \SI{200}{\nano\meter}, above a magnetic layer extending across \SI{-100}{\nano\meter} $<z<$ \SI{100}{\nano\meter}.
Because the external magnetic field is preferably applied along the nanowire, while a transverse gradient is required, we rely on a uniform magnetization $\vect{m}$ along $+x$.
As we now consider dots located above the magnetic system, we use $Q = \int B_{y/z}(x) \norm{\psi_{\rm{L}}(x)}^2 {\rm{d}}x  - \int B_{y/z}(x) \norm{\psi_{\rm{R}}(x)}^2 {\rm{d}}x $, where $L$ and $R$ are exchanged, compared to the above.
To ensure practical shapes, a constraint was added that the magnet is homogeneous across $z$.
Thus, this optimization consists in deciding the presence or absence of magnet in cells of the full thickness at all $u,v$ positions.
The resulting optimal nanomagnets for $B_{y}$ and $B_{z}$ difference are shown in Fig.\ \ref{fig:idealcase2}(a--d) and (e--h), respectively.
Similar to before, the shapes are explained by the ideal distributions, but are now convolved with the finite extension of the dots.
The negligible differences in the present example are highlighted by Fig.\ \ref{fig:idealcase2}(a,e), which display the corresponding magnet boundaries with and without the finite size of dots for either $B_{y}$ or $B_{z}$.
This shows that in standard cases, the finite extension of the dots can be safely neglected in what concerns the design of the nanomagnets.
The linear profiles of $\vect{B}(x)$ obtained along the nanowire are displayed in Fig.\ \ref{fig:idealcase2}(c,g).
The relevant part of these profiles, determining the energy levels in the dots, is the one overlapping with $\norm{\psi_p(\vect{r})}^2$ from each dot.
An important question beyond the shape of the nanomagnets is their expected footprint on a chip.
It is thus desirable to know which parts of the nanomagnets are crucial to generate the gradient, and which parts can be disregarded.
The importance of the contributions from each part of the nanomagnets to $Q$ are shown in Fig.\ \ref{fig:idealcase2}(d,h). 
It confirms that the regions contributing the most to $Q$ (or $\alpha_{\rm{as}}$) are strongly localized near the dots.
This highlights the critical importance of mastering the nanofabrication processes to achieve well-defined geometrical shapes at these internal edges of the nanomagnets.
%

%
\section{Accurate modeling with micromagnetic optimization}\label{sec:micromag:mumax}
%
%

\begin{figure*}
\centering

\includegraphics[clip]{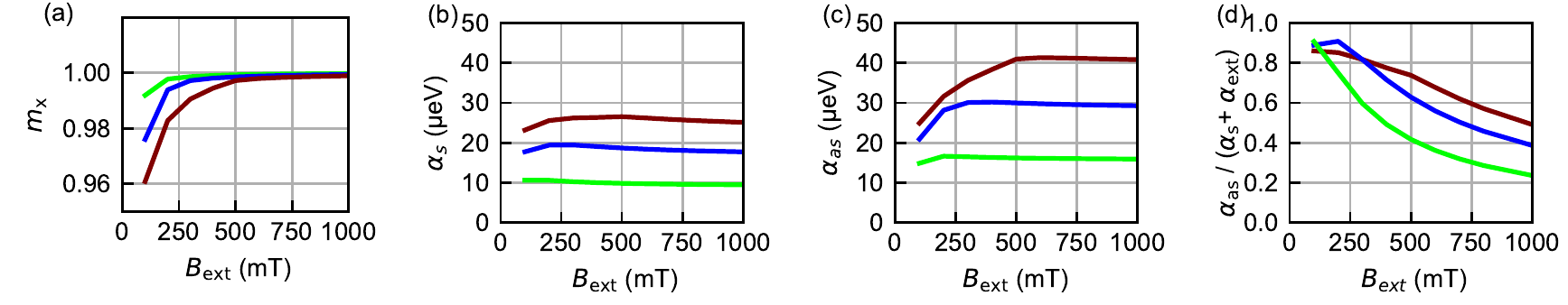}

\caption{Results of the micromagnetic simulations of an optimized shape for an inhomogeneous $B_{\rm{z}}$ component, with external field $B_{\rm{ext}}$ and magnetization along $x$. As a function of external magnetic field, (a) magnetization $m_{\rm{x}}$, (b) symmetric ($\alpha_{s}$), (c) antisymmetric ($\alpha_{as}$) coupling constants due to the nanomagnets, and (c) the ratio between the antisymmetric and the total symmetric field $\alpha_{as}/(\alpha_{s}+\alpha_{\rm{ext}})$. The different colors correspond to CoFe (red), Co (blue) and NiFe (green).}
\label{fig:mumax}
\end{figure*}

The stray fields exhibited above are only valid under the condition of magnetic saturation of the nanomagnets.
Even though an external field is usually applied to orientate the magnetic system, the internal magnetic interactions tend to disfavor this uniform magnetization and may lead to the formation of inhomogeneous domains inside the structures.
For our previously optimized designs, we thus aim to find what minimum external field is required to ensure magnetic saturation, and to understand by how much the stray field gradients are affected when this condition is not met.
Having found optimal shapes for different uniform configurations above, we now consider an accurate modeling of all the magnetostatic interactions that will shape the internal magnetization of the nanomagnets.
Micromagnetic simulations are able to capture the differences between a uniform magnetization and a spatially-varying magnetization, affected by finite-size effects and internal interactions (direct exchange and dipolar terms).
To illustrate our approach, micromagnetic simulations are performed with the Mumax$^3$ software \cite{Vansteenkiste2014} for a magnet of case (j) of Fig.\ \ref{fig:munif} above. 
The simulated magnetic system has dimensions of 7000 $\times$ 1000 $\times$ \SI{200}{nm\cubed} and a cell size of 4 $\times$ 4 $\times$ \SI{4}{nm\cubed}, see Supplemental Material S4 and S5 for details.
Following magnetic saturation with an external field of \SI{1}{\tesla}, the stable magnetic configurations under different values of the external field $\vect{B}_{\rm{ext}}$ are obtained by decreasing steps of \SI{100}{\milli\tesla}.
Considering the saturation magnetization $M_{\rm{S}}$ of the magnetic material is important for the present micromagnetic modeling, as it sets the energy scale of the dipolar interactions, entering as a quadratic term into the domains formation energy.
While a larger $M_{\rm{S}}$ directly favors larger stray fields and gradients, it also facilitates the formation of magnetic domains in the nanomagnets and thus require a larger external field to reach saturation.
This external field adds up as a term $\alpha_{\rm{ext}}=2\mu_{\rm{B}}B_{\rm{ext}}$ to $\alpha_{\rm{s}}$ from the symmetric part of the stray fields in forming the dots energy levels.
Defining the inhomogeneity of the fields acting on the dots as $\alpha_{\rm{as}}/(\alpha_{\rm{s}}+\alpha_{\rm{ext}})$, this figure of merit value reduces when a large external field is employed.
To optimize the inhomogeneity, it is thus required to identify the pairs of $M_{\rm{S}}$ and $\vect{B}_{\rm{ext}}$ values providing the strongest $\alpha_{\rm{as}}/(\alpha_{\rm{s}}+\alpha_{\rm{ext}})$.
Standard materials for nanomagnet fabrication include FeCo, Co or NiFe, with values of $M_{\rm{S}}$ around \SI{2000}{\kilo\ampere\per\meter}, \SI{1450}{\kilo\ampere\per\meter} and \SI{800}{\kilo\ampere\per\meter} at $<$ \SI{4}{\kelvin} temperatures, respectively.
These values of $M_{\rm{S}}$ can be further adjusted down by alloying the ferromagnet with a non-magnetic element, in order to reach a precise $M_{\rm{S}}$ requirement \cite{Eyrich2014}.
We present in Fig.\ \ref{fig:mumax}(a) the evolution of the magnetization $m_{\rm{x}}$, in Fig.\ \ref{fig:mumax}(b,c) that of the coupling terms $\alpha_{\rm{s}}$ and $\alpha_{\rm{as}}$, and in Fig.\ \ref{fig:mumax}(d) that of the inhomogeneity ratio $\alpha_{\rm{as}}/(\alpha_{\rm{s}}+\alpha_{\rm{ext}})$, as a function of the external field for these materials with different $M_{\rm{S}}$.
It appears that within the usual range of employed external fields, the choice of magnetic material is a key parameter of the system, as saturation is reached at about \SI{200}{\milli\tesla} for NiFe, \SI{300}{\milli\tesla} for Co and \SI{600}{\milli\tesla} for CoFe.
To ensure reproducible fields at given conditions, it is necessary to ensure magnetic saturation.
Depending on the admissible range of external fields in the system, and on whether a large inhomogeneity or a large nominal gradient of field is required, a particular pair of $M_{\rm{S}}$ and $\vect{B}_{\rm{ext}}$ values might be more suitable.
As is highlighted by Fig.\ \ref{fig:mumax}(d), some of the designs presented above, including (j) of Fig.\ \ref{fig:munif}, can reach a regime where $\alpha_{\rm{as}}\sim(\alpha_{\rm{s}}+\alpha_{\rm{ext}})$.
%
%
\section{Conclusion}\label{sec:conclusion}
Despite its rather straightforward concepts, the present treatment provides a valuable and simple solution to the design of field gradients with nanomagnets.
These are also compatible with rotated designs that allow for the differential tuning of the Zeeman energy splitting in different DQDs with tilted external fields \cite{Borjans2019}, or with quantum dot arrays \cite{Nakamura2022}.
Extension of these ideas to the generation of cycloidal periodic fields, also of interest to shape synthetic spin--orbit coupling in quantum dots, nanowires and two-dimensional electron gases \cite{Zhou2019,Mohanta2019,Kornich2020,Turcotte2020}, can in principle be realized by considering a periodic space for this optimization.
The case of a perpendicularly magnetized system hosting domains has also been briefly presented. 
Taking into account that a magnetic multilayer, such as Pt/Co or Co/Ni, provides the required out-of-plane anisotropy at the cost of a reduced $M_{\rm{S}}$ (due to the dilution of the magnetic moments inside a superstructure also containing less or non-magnetic materials), it is left to future investigation to distinguish whether this can be beneficial overall.
The present approach is appealing for many material platforms hosting spin qubits: two-dimensional, such as silicon, germanium, graphene \cite{Burkard2020}, or one-dimensional, such as semiconducting nanowires and carbon nanotubes \cite{Viennot2015,Cubaynes2019}, etc., as it does not rely on an intrinsic, defect-dependent spin--orbit interaction but on an extrinsic spin--orbit coupling tunable by the arrangement of nanomagnet shapes.
For each of them, the achievable gradients are largely influenced by maximal interdot distance and minimal dot-to-nanomagnet separation.
For nanowire spin qubits, this also provides an alternative to the use of ferromagnetic contacts \cite{Cottet2010,Viennot2015}.
These results will contribute to efforts for applying gate operations to distant spin qubits through their coupling to microwave radiation in high-impedance resonators at low powers ($\sim$ single photon), in the framework of circuit quantum electrodynamics \cite{Burkard2020}.
By comparing the benefits of strong spin--charge hybridization to decoherence contributions, it appears that a desirable regime for these operations is found around $\alpha_{\rm{as}}\sim\alpha_{\rm{s}}$.
This situation, where the strength of the antisymmetric component reaches the strength of the symmetric component of the stray fields, corresponds to a particularly high inhomogneity.
There, the effective fields acting over each dot point in directions differing by an angle of almost \SI{90}{deg}.
We find here some optimal shapes for nanomagnets that achieve this regime and that are feasible with standard nanofabrication techniques, enabling different field gradient orientations.
They offer an improvement of the magnetic field difference between dots by a factor at least two and up to ten, with respect to some earlier designs.
As the energy associated to these fields determines the addressability of the spin qubits, improving the gradient will help moving beyond the perturbative regime of operation that has been commonly employed.
With the results presented above, high-fidelity gate operations could be targeted in spin-qubit devices, while the optimized designs can also be applied to other research fields employing nanomagnets.

\begin{acknowledgments}
We thank A.\ Thiaville for fruitful conversations and critical reading of the manuscript.
\end{acknowledgments}

%

%

%
\end{document}